\documentclass[12pt,preprint]{aastex}
\usepackage{graphicx}  
\usepackage{dcolumn}   
\usepackage{bm}        
\usepackage{amssymb}   
\usepackage{natbib}
\begin{document}

\title{The signature of an anisotropic distribution of gamma-ray bursts}

\author{Lixiong Gan\altaffilmark{1}, Yuan-Chuan Zou\altaffilmark{1}, Zi-Gao Dai\altaffilmark{2}}
\altaffiltext{1}{School of Physics, Huazhong University of Science and Technology, Wuhan 430074, China; ganli2012@gmail.com(LXG) zouyc@hust.edu.cn(YCZ)}
\altaffiltext{1}{School of Astronomy and Space Science, Nanjing University, Nanjing 210093, China; dzg@nju.edu.cn(ZGD)}

\begin{abstract}
Anomalies of the cosmic microwave background (CMB) maps have been widely acquainted nowadays from the Wilkinson Microwave Anisotropy Probe (WMAP) satellite to the Planck satellite. One of the anomalies is a multipole alignment from $l=2$ to $l=5$. In our work, we investigate the angular distribution of gamma-ray bursts (GRBs) to find whether there is the same anomaly of GRB as CMB. We perform spherical harmonics expansion on GRB samples to derive coefficients of a few first terms of the expansion terms and find that there is rough multipole alignment from $l=2$ to $l=4$ while the dipole and $l=5$ multipole is in a distant direction, and that the quadrupole is obviously planar and the other ones are normal.
\end{abstract}

\maketitle

\section{Introduction}

The cosmic microwave background (CMB) is an observational phenomenon that reflects the profound properties of our universe. A decade ago, some anomalies \citep{2004PhRvD..69f3516D} have been discovered by analysis of CMB results from the Wilkinson Microwave Anisotropy Probe (WMAP). The anomalies include: 1) The cosmic quadrupole on its own is anomalous at the 1-in-20 level by being low; 2) The cosmic octopole on its own is anomalous at the 1-in-20 level by being very planar; 3) The alignment between the quadrupole and octopole is anomalous at the 1-in-60 level. Further work has been done, focusing on these anomalies \citep{2006MNRAS.367...79C,2004PhRvL..93v1301S,2004IJMPD..13.1857R}. And anomalies actually extend to the higher multipoles $l$=2--5 \citep{2005PhRvL..95g1301L}. Recently, a much more precise CMB map from the Planck satellite provides access to a detailed analysis, and finally confirmed the anomalies with higher confidence \citep{2013arXiv1303.5062P,2013arXiv1303.5083P}. Until now anomalies from CMB remain puzzled.

Besides CMB, there are some similar phenomena found in other observations. Through analysis of quasar spectroscopy, the fine structure constant is found distributed anisotropically whose dataset fits a spatial dipole in the direction right ascension
$17.5\pm 0.9$ hours, declination $-58 \pm 9$ degrees \citep{2011PhRvL.107s1101W}. Through analysis of supernovae, there is also a preferred direction of the acceleration of the universe which is located roughly at $(l,b) \sim (130,0)$ \citep{2012JCAP...02..004C,2013MNRAS.tmp.2732Y}.

Inspired by the intriguing anomalies above, we pay attention to GRBs. Contrary to CMB, GRBs are violent common events in the universe, and the detection rate is a few events per day by some instruments. GRBs are of high redshifts among most observable objects in the universe, thus containing much information about our universe. One of our interests is the sky distribution of GRBs, which has been studied by some works \citep{1998A&A...339....1B, 2001grba.conf...56C, 2009AIPC.1133..483M, 2009BaltA..18..293M, 2010AIPC.1279..457V, 2014arXiv1406.6480K}. It has been found that short and intermediate GRBs distribute anisotropically while long GRBs approximately isotropically.  Furthermore, \citet{2000AstL...26..672T} proposed a simple method to derive quadrupole of the GRB distribution and found no obvious quadrupole.

In this paper, we use an approach similar to the CMB analysis, spherical harmonic expansions, to obtain the coefficients of the expansion terms and then perform the same steps as \citet{2004PhRvD..69f3516D} to check whether there are anomalies in the angular distribution of GRBs. Unlike the CMB, the angular distribution of GRBs can be started from the dipole. In section 2, we introduce an approach to derive spherical harmonic coefficients, and simulate a sample of some specific distribution mode to test the approach. In section 3, we apply the method in section 2 to the GRB sample. Conlusion and discussion are made in section 4.

\section{Spherical Harmonics Transform}

A continuous function on a spherical surface can be developed into spherical harmonic coefficients,

\begin{equation} \label{eq1}
f(\theta,\varphi)=\sum\limits_{l}\sum\limits_{m}a_{lm}Y_{lm}(\theta,\varphi),
\end{equation}
where
\begin{equation} \label{eq2}
a_{lm}=\int\limits_{\theta}\int\limits_{\varphi}f(\theta,\varphi)Y_{lm}(\theta,\varphi)^{*}\sin{\theta}{\rm d}\theta{\rm d}\varphi.
\end{equation}

The spherical harmonics $Y_{lm}(\theta,\varphi)$ are normalized as
\begin{equation} \label{eq3}
\int\limits_{\theta}\int\limits_{\varphi}Y_{lm}(\theta,\varphi)Y_{l'm'}(\theta,\varphi)^{*}\sin{\theta}{\rm d}\theta{\rm d}\varphi=\delta_{ll'}\delta_{mm'}.
\end{equation}

The above method is aimed at transforming continuous distributions. However, when processing practical situations we are almost faced with discrete distributions. Here, we use a little trick to tackle the problem.

In a discrete distribution, a function is only defined in separated points, so we cannot derive coefficients by integral at all. We naturally transform integral to summation,
\begin{equation} \label{eq4}
a_{lm}=\sum\limits_{i}f(\varsigma _{i},\xi _{i})Y_{lm}(\varsigma _{i},\xi _{i})^{*}\Delta S(\varsigma _{i},\xi _{i}),
\end{equation}
where we adopt curved coordinates $(\varsigma _{i},\xi _{i})$, and subscript $i$ denotes different coordinates in different areas. $\Delta S(\varsigma _{i},\xi _{i})$ is the area where $(\varsigma _{i},\xi _{i})$ lies. If $\Delta S(\varsigma _{i},\xi _{i})$ is infinitesimal, the summation is transformed to integral. Furthermore, we choose mean values of $f(\varsigma _{i},\xi _{i})$ in the area $\Delta S(\varsigma _{i},\xi _{i})$ as an approximation in our case,
\begin{equation} \label{eq5}
f(\varsigma _{i},\xi _{i})=\frac{n_{i}}{\Delta S(\varsigma _{i},\xi _{i})},
\end{equation}
where $n_{i}$ is the number of GRBs in the area $\Delta S(\varsigma _{i},\xi _{i})$. The way that we divide the spherical surface is arbitrary, and we can always make $\Delta S(\varsigma _{i},\xi _{i})$ as small as possible so that we have not more than one GRB in an arbitrary area. In this way, the expression of coefficients is written as
\begin{eqnarray} \label{eq6}
a_{lm} &=& \sum\limits_{i}\frac{1}{\Delta S(\varsigma _{i},\xi _{i})}Y_{lm}(\varsigma _{i},\xi _{i})^{*}\Delta S(\varsigma _{i},\xi _{i}) \nonumber \\
&=& \sum\limits_{i}Y_{lm}(\varsigma _{i},\xi _{i})^{*}=\sum\limits_{i}Y_{lm}(\theta _{i},\varphi _{i})^{*},
\end{eqnarray}
where curved coordinates are converted back to spherical coordinates, and subscript $i$ denotes different GRBs.


We combine the method described above and the method in \citet{2004PhRvD..69f3516D} to search for multipoles. Firstly, we adopt specific distributions to test our method, i.e. distributions like spherical harmonics $Y_{lm} (\theta, \varphi )$. For instance, if a distribution is exactly shaped like $Y_{20} (\theta, \varphi )$, then we will find a quadrupole lying on the equatorial plane. If a distribution is exactly shaped like $Y_{33} (\theta ,\varphi )$, we will find octopole pointing to the north pole.

We simulated 50000 points uniformly distributed on the spherical surface and attach them to $Y_{20} (\theta ,\varphi )$ so that $Y_{20} (\theta ,\varphi )$ is defined uniformly. Then we attain the desired function which is discrete.  Using the method of the previous method, the five coefficients $a_{2m} $ are derived. Rotate the coefficients to search for quadrupole and finally find it in the direction of $(\theta ,\varphi )\sim (90^{\circ},232.5^{\circ})$ without surprise. The unrotated and rotated real-valued coefficients are listed in Table 1. Besides $Y_{20} (\theta, \varphi )$, we can as well find the pole of any other distribution mode in the right direction.

\begin{table} \label{table1}
\caption{The unrotated and rotated real-valued quadrupole coefficients.}
\begin{tabular}{|p{0.9in}|p{0.9in}|p{0.9in}|} \hline
coefficients & unrotated & rotated \\ \hline
$a_{2-2} $ & 4.53 & -7.68 \\ \hline
$a_{2-1} $ & 3.32 & 4.35 \\ \hline
$a_{20} $ & 3986.06 & -1966.02 \\ \hline
$a_{21} $ & 7.68 & -3.32 \\ \hline
$a_{22} $ & -31.19 & 3467.63 \\ \hline
\end{tabular}
\end{table}

\section{Results}
With the methods described above, we are able to cope with the angular distribution of GRBs. Here, we adopt a sample of 3899 GRBs where 1236 GRBs are from http://www.mpe.mpg.de/ \~{}jcg/grbgen.html (this website is updated daily and we select GRBs from the oldest one to GRB 131014A. We have removed GRBs involved in the following sample.) and 2663 are from BATSE 4B Catalog \citep{1999ApJS..122..465P} (the website http://www.batse.msfc.nasa.gov/batse/grb /catalog/current/tables/basic\_table.txt). Both are in Galactic coordinates.

Upon figuring out our desired coefficients of the spherical harmonics expansion, we quickly get the preferred axes from $\stackrel{\frown}{n}_{1} $ to $\stackrel{\frown}{n}_{5} $ of different multipoles respectively, as follows,
\begin{equation} \label{eq9}
\begin{array}{l} {\stackrel{\frown}{n}_{1} =(-0.1198,-0.3949,0.9109)} \\ {\stackrel{\frown}{n}_{2} =(0.4811,-0.2778,0.8315)} \\ {\stackrel{\frown}{n}_{3} =(0.6648,-0.3012,0.6836)} \\ {\stackrel{\frown}{n}_{4} =(0.7778,-0.4157,0.4714)} \\ {\stackrel{\frown}{n}_{5} =(-0.7852,0.5626,0.2588)} \end{array} .
\end{equation}
Converted to Galactic coordinates, they are roughly in the direction of
\begin{equation} \label{eq10}
\begin{array}{l} {(b,l)_{1} \sim (22^{\circ } ,73^{\circ } )} \\ {(b,l)_{2} \sim (33^{\circ } ,150^{\circ } )} \\ {(b,l)_{3} \sim (47^{\circ } ,156^{\circ } )} \\ {(b,l)_{4} \sim (62^{\circ } ,152^{\circ } )} \\ {(b,l)_{5} \sim (75^{\circ } ,-36^{\circ } )} \end{array} .
\end{equation}
It is easy to see that the second to the fourth vectors have some kind of alignment while the last and the first oneis in a totally different direction. Figure 1 demonstrates the map from $l=1$ to $l=5$ respectively, where deep green denotes positive values and light green denotes negative ones. From the dipole and quadrupole map, we find they are more planar than others, and the corresponding direction is perpendicular to the plane where deep and light green varies from each other. Figure 2 shows the distributions of the angular dispersion on the spherical surface, where larger values are donated in the deeper blue. Clearly, in each map the two ends of the direction of multipoles lie in the two deep blue areas, which can be regarded as possible errors of each direction respectively.

The correlations of the second to the fourth vectors are
\begin{equation} \label{eq11}
\begin{array}{l} {\theta _{23} =13.6^{\circ } } \\ {\theta _{24} =28.2^{\circ } } \\ {\theta _{34} =15.3^{\circ } } \end{array}.
\end{equation}
Although these angles are not small, they are yet not large enough to be ignored, especially the angle between quadrupole and octopole. Comparing with the axes of CMB multipoles \citep{2004PhRvD..69f3516D},
\begin{equation} \label{eq12}
\begin{array}{l} {\stackrel{\frown}{n}_{2} =(-0.1145,-0.5265,0.8424)} \\ {\stackrel{\frown}{n}_{3} =(-0.2578,-0.4207,0.8698)} \end{array} ,
\end{equation}
we find that the third component of the above two factors agrees well with that of $\stackrel{\frown}{n}_{2} $ in equation (\ref{eq9}). That means the $\theta $ directions of axes in both cases are similar. However, the first and the second component of each vector in equation (\ref{eq12}) are inconsistent with that in equation (\ref{eq9}). Thus we conclude that the $\varphi $ directions in both cases have no common point.

Besides, in GRB distribution $\stackrel{\frown}{n}_{5} $ differs much from the first three vectors and has no alignment with them. In detail, $\stackrel{\frown}{n}_{5} $ has a separation of about 71.4 degrees with $\stackrel{\frown}{n}_{2} $. In contrary to GRB, the $l=5$ multipole for the CMB map is aligned with $(b,l)\sim (50^{\circ } ,-91^{\circ } )$ \citep{2004PhRvD..69f3516D}.

Apart from $l=5$, the dipole also points to a different direction, which misaligned with others. The angles between them are as follows:  \begin{equation} \label{eq15}
\begin{array}{l} {\theta _{12} =35^{\circ } } \\ {\theta _{13} =48^{\circ } } \\ {\theta _{14} =60^{\circ } } \\ {\theta _{15} =83^{\circ } }\end{array}.
\end{equation}

Furthermore, we calculate a statistical parameter $t_{l} $ that describes how planar each multipole is. The expression of $t_{l} $ is \citep{2004PhRvD..69f3516D}
\begin{equation} \label{eq13}
t_{l} =\mathop{\max }\limits_{\stackrel{\frown}{n}} \frac{|a_{l-l} (\stackrel{\frown}{n})|^{2} +|a_{ll} (\stackrel{\frown}{n})|^{2} }{\sum _{m=-l}^{l}|a_{lm} (\stackrel{\frown}{n})|^{2}  } .
\end{equation}
The coefficients $a(\stackrel{\frown}{n})$ above are all rotated to the preferred frame where the angular momentum dispersion is maximized. The higher $t_{l} $ is, more dominant the $|m|=l$ mode is, therefore the $l$ multipole is more obvious. $t_{l} $ for GRB distribution are listed below:
\begin{equation} \label{eq14}
\begin{array}{l} {t_{1} =100\% } \\ {t_{2} =91\% } \\ {t_{3} =68\% } \\ {t_{4} =43\% } \\ {t_{5} =18\% } \end{array} .
\end{equation}
Dipole is obviously planar.We can also find quadrupole planar in the quadrupole map. Beside quadrupole, there is no anomaly in other multipoles. For comparison, $t$ for octopole of the CMB map is 94\% \citep{2004PhRvD..69f3516D}.

\section{Discussion}

Taking the angular position of 3899 GRBs, we have calculated axes for the multipoles from $l=1$ to $l=5$, and found some anomalies, i.e., the axes of $l=2,3,4$ are aligned with angles $\theta_{23} =13.6^{\circ },  \theta_{24} =28.2^{\circ }, \theta _{34} =15.3^{\circ }$ respectively, while the axes of $l=1$ and $l=5$ are directed far away. The aligned axes of $l=2,3,4$ are not aligned with the CMB axis \citep{2004PhRvD..69f3516D}. Notice there also exist prefered directions for the fine structure \citep{2011PhRvL.107s1101W}, and supernovae \citep{2012JCAP...02..004C,2013MNRAS.tmp.2732Y}, but no any two of them  are aligned.

We should remember our analytical method is not so robust as expected. The way that we expand the GRB distribution is an approximate approach, and has an effect on the coefficients. Specifically, equation (\ref{eq5}) means that we adopt a function of constant GRB number density $f(\varsigma _{i} ,\xi _{i} )$ instead of GRB number in area $\Delta S(\varsigma _{i} ,\xi _{i} )$, i.e. we assign each point on spherical surface a value derived from nearby GRBs. Therefore, what we did is to transit the real GRB map to a similar map where all points on the surface have definition. We don't know errors from the transition and whether it will swing largely axes of multipoles.

Errors of GRB localization are not taken into consideration. As we know, many GRBs do not have definite localization and some may range over one degree. As a result, even the real GRB map is not as accurate as the CMB map. We should also notice that GRBs are detected by several instruments and each has different thresholds, which may induce the somewhat selection effect.

Finally, if the effect mentioned above is small enough, we are glad to see a potential correlation between the GRB map and the CMB map. Two maps originate from different physical scenarios. CMB originates from transmission of photons from last scattering surface to observers and anisotropy of the CMB is due to fluctuation of the early universe. On the other hand, anisotropy of GRBs distribution may be the consequence of fluctuation of mass density during primordial universe or sequence of the universe evolution. Whether they have a common point or correlate with each other is ambiguous. But, if we do indeed find some relationship between the axis of evil of GRBs and CMB, we obtain some clues to reveal the structure of our universe behind the uncanny phenomenon.

YCZ thanks the helpful discussion with Tsvi Piran, Dingxiong Wang, Bing Zhang, Biping Gong, Qingwen Wu and Weihua Lei. This work is supported by the National Basic Research Program of China (973 Program, Grant No. 2014CB845800), the National Natural Science Foundation of China (Grants No. U1231101 and 11033002) and the Chinese-Israeli Joint Research Project (Grant No. 11361140349).

\def\aap{A\&A }
\def\aaps{A\&AS}
\def\apjl{ApJL }
\def\aj{AJ }
\def\apj{ApJ }
\def\apjs{ApJS }
\def\araa{ARAA }
\def\jcap{JCAP}
\def\mnras{MNRAS }
\def\nat{Nature }
\def\pasp{PASP }
\def\physrep{Physics Report}
\def\spie{SPIE }

\begin{figure}
 \includegraphics[width=0.5\textwidth]{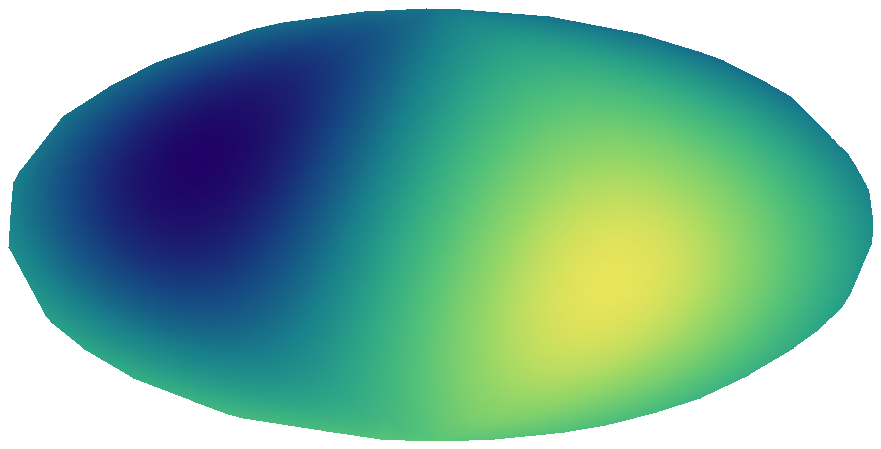}
 \includegraphics[width=0.5\textwidth]{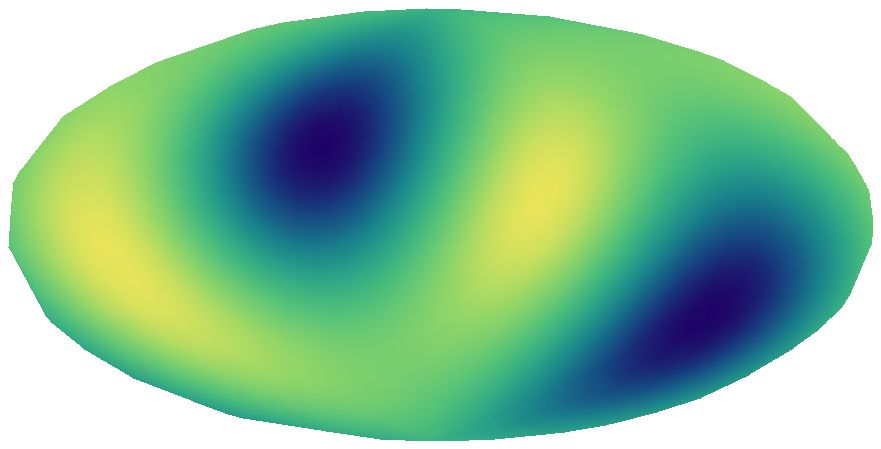}
 \includegraphics[width=0.5\textwidth]{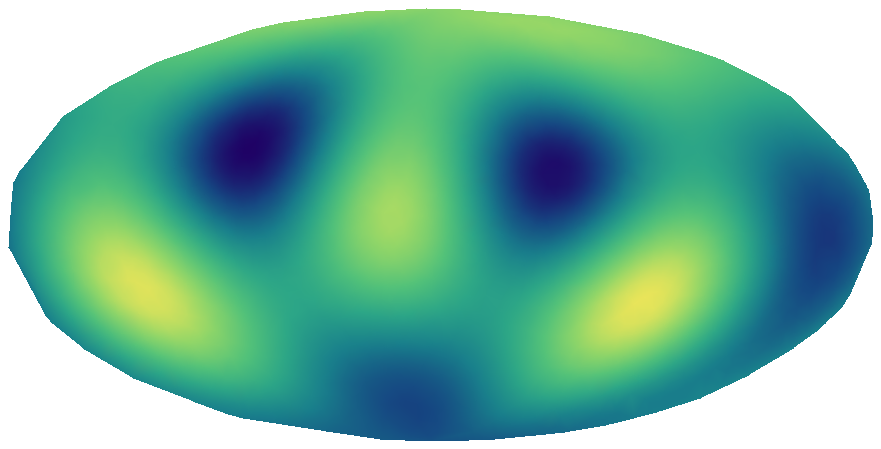}
 \includegraphics[width=0.5\textwidth]{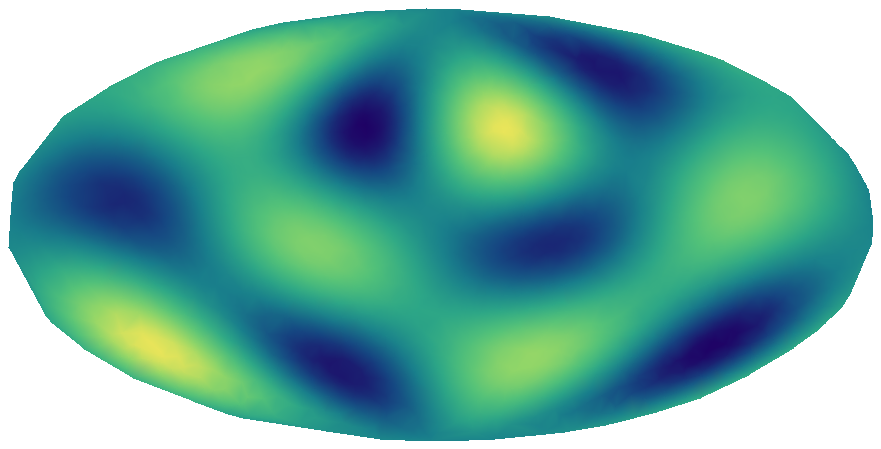}
 \includegraphics[width=0.5\textwidth]{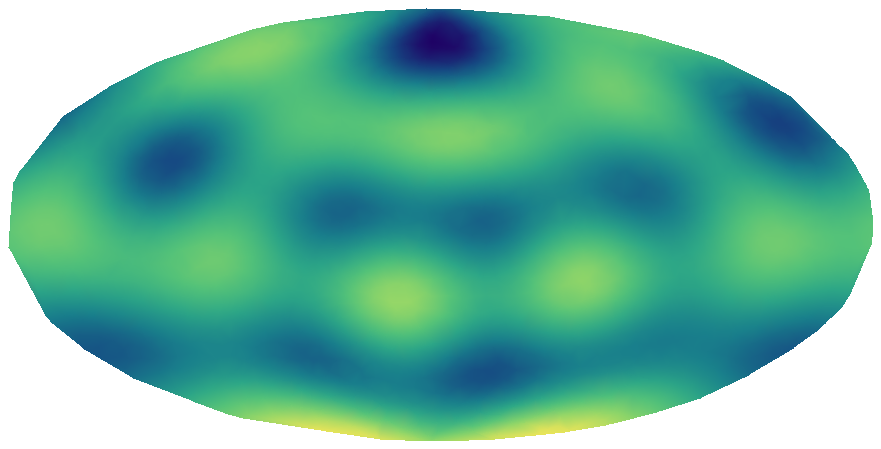}
 \caption{From $l=1$ map to $l=5$ map. Deep green denotes positive values and light green denotes negative ones. }
 \label{fig-spectra}
\end{figure}

\begin{figure}
 \includegraphics[width=0.5\textwidth]{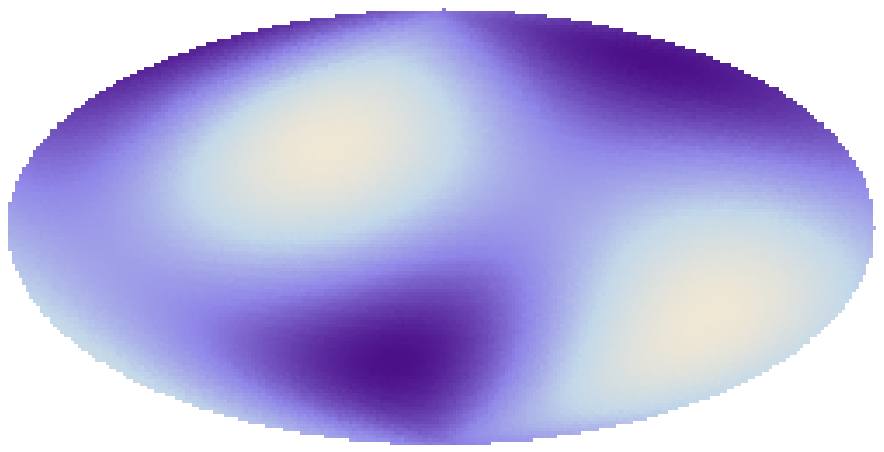}
 \includegraphics[width=0.5\textwidth]{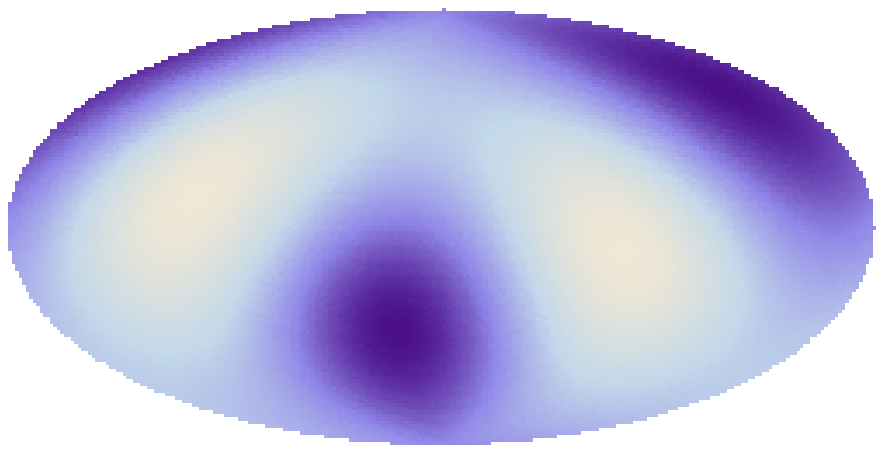}
 \caption{Distributions of angular dispersion in the spherical surface for $l=2$ and $l=3$, where larger values are colored in deeper blue.}
 \label{fig-spectra}
\end{figure}


\begin{thebibliography}{}
\expandafter\ifx\csname natexlab\endcsname\relax\def\natexlab#1{#1}\fi

\bibitem[{{Balazs} {et~al.}(1998){Balazs}, {Meszaros}, \&
  {Horvath}}]{1998A&A...339....1B}
{Balazs}, L.~G., {Meszaros}, A., \& {Horvath}, I. 1998, \aap, 339, 1

\bibitem[{{Cai} \& {Tuo}(2012)}]{2012JCAP...02..004C}
{Cai}, R.-G., \& {Tuo}, Z.-L. 2012, \jcap, 2, 4

\bibitem[{{Cline} {et~al.}(2001){Cline}, {Matthey}, \&
  {Otwinowski}}]{2001grba.conf...56C}
{Cline}, D.~B., {Matthey}, C., \& {Otwinowski}, S. 2001, in Gamma-ray Bursts in
  the Afterglow Era, ed. E.~{Costa}, F.~{Frontera}, \& J.~{Hjorth}, 56

\bibitem[{{Copi} {et~al.}(2006){Copi}, {Huterer}, {Schwarz}, \&
  {Starkman}}]{2006MNRAS.367...79C}
{Copi}, C.~J., {Huterer}, D., {Schwarz}, D.~J., \& {Starkman}, G.~D. 2006,
  \mnras, 367, 79

\bibitem[{{de Oliveira-Costa} {et~al.}(2004){de Oliveira-Costa}, {Tegmark},
  {Zaldarriaga}, \& {Hamilton}}]{2004PhRvD..69f3516D}
{de Oliveira-Costa}, A., {Tegmark}, M., {Zaldarriaga}, M., \& {Hamilton}, A.
  2004, \prd, 69, 063516

\bibitem[{{Khabibullina} {et~al.}(2014){Khabibullina}, {Verkhodanov}, \&
  {Sokolov}}]{2014arXiv1406.6480K}
{Khabibullina}, M.~L., {Verkhodanov}, O.~V., \& {Sokolov}, V.~V. 2014, ArXiv
  e-prints, arXiv:1406.6480

\bibitem[{{Land} \& {Magueijo}(2005)}]{2005PhRvL..95g1301L}
{Land}, K., \& {Magueijo}, J. 2005, Physical Review Letters, 95, 071301

\bibitem[{{Meszaros} {et~al.}(2009){Meszaros}, {Balazs}, {Bagoly}, \&
  {Veres}}]{2009AIPC.1133..483M}
{Meszaros}, A., {Balazs}, L.~G., {Bagoly}, Z., \& {Veres}, P. 2009, in American
  Institute of Physics Conference Series, Vol. 1133, American Institute of
  Physics Conference Series, ed. C.~{Meegan}, C.~{Kouveliotou}, \&
  N.~{Gehrels}, 483--485

\bibitem[{{M{\'e}sz{\'a}ros} {et~al.}(2009){M{\'e}sz{\'a}ros}, {Bal{\'a}zs},
  {Bagoly}, \& {Veres}}]{2009BaltA..18..293M}
{M{\'e}sz{\'a}ros}, A., {Bal{\'a}zs}, L.~G., {Bagoly}, Z., \& {Veres}, P. 2009,
  Baltic Astronomy, 18, 293

\bibitem[{{Paciesas} {et~al.}(1999){Paciesas}, {Meegan}, {Pendleton}, {Briggs},
  {Kouveliotou}, {Koshut}, {Lestrade}, {McCollough}, {Brainerd}, {Hakkila},
  {Henze}, {Preece}, {Connaughton}, {Kippen}, {Mallozzi}, {Fishman},
  {Richardson}, \& {Sahi}}]{1999ApJS..122..465P}
{Paciesas}, W.~S., {Meegan}, C.~A., {Pendleton}, G.~N., {et~al.} 1999, \apjs,
  122, 465

\bibitem[{{Planck Collaboration} {et~al.}(2013{\natexlab{a}}){Planck
  Collaboration}, {Ade}, {Aghanim}, {Armitage-Caplan}, {Arnaud}, {Ashdown},
  {Atrio-Barandela}, {Aumont}, {Baccigalupi}, {Banday}, \&
  et~al.}]{2013arXiv1303.5062P}
{Planck Collaboration}, {Ade}, P.~A.~R., {Aghanim}, N., {et~al.}
  2013{\natexlab{a}}, ArXiv e-prints, arXiv:1303.5062

\bibitem[{{Planck Collaboration} {et~al.}(2013{\natexlab{b}}){Planck
  Collaboration}, {Ade}, {Aghanim}, {Armitage-Caplan}, {Arnaud}, {Ashdown},
  {Atrio-Barandela}, {Aumont}, {Baccigalupi}, {Banday}, \&
  et~al.}]{2013arXiv1303.5083P}
---. 2013{\natexlab{b}}, ArXiv e-prints, arXiv:1303.5083

\bibitem[{{Ralston} \& {Jain}(2004)}]{2004IJMPD..13.1857R}
{Ralston}, J.~P., \& {Jain}, P. 2004, International Journal of Modern Physics
  D, 13, 1857

\bibitem[{{Schwarz} {et~al.}(2004){Schwarz}, {Starkman}, {Huterer}, \&
  {Copi}}]{2004PhRvL..93v1301S}
{Schwarz}, D.~J., {Starkman}, G.~D., {Huterer}, D., \& {Copi}, C.~J. 2004,
  Physical Review Letters, 93, 221301

\bibitem[{{Tikhomirova} \& {Stern}(2000)}]{2000AstL...26..672T}
{Tikhomirova}, Y.~Y., \& {Stern}, B.~E. 2000, Astronomy Letters, 26, 672

\bibitem[{{Veres} {et~al.}(2010){Veres}, {Bagoly}, {Horv{\'a}th}, {Bal{\'a}zs},
  {M{\'e}sz{\'a}ros}, \& {Kelemen}}]{2010AIPC.1279..457V}
{Veres}, P., {Bagoly}, Z., {Horv{\'a}th}, I., {et~al.} 2010, in American
  Institute of Physics Conference Series, Vol. 1279, American Institute of
  Physics Conference Series, ed. N.~{Kawai} \& S.~{Nagataki}, 457--459

\bibitem[{{Webb} {et~al.}(2011){Webb}, {King}, {Murphy}, {Flambaum},
  {Carswell}, \& {Bainbridge}}]{2011PhRvL.107s1101W}
{Webb}, J.~K., {King}, J.~A., {Murphy}, M.~T., {et~al.} 2011, Physical Review
  Letters, 107, 191101

\bibitem[{{Yang} {et~al.}(2013){Yang}, {Wang}, \& {Chu}}]{2013MNRAS.tmp.2732Y}
{Yang}, X., {Wang}, F.~Y., \& {Chu}, Z. 2013, \mnras, arXiv:1310.5211

\end{thebibliography}
\end{document}